\documentclass[12pt]{article}
\usepackage{putex}
\usepackage{graphicx}

\begin{document}

\preprint{PUPT-2316}

\institution{PU}{${}^1$Joseph Henry Laboratories, Princeton University, Princeton, NJ 08544, USA}
\institution{Barcelona}{${}^2$Universitat Polit\`ecnica de Catalunya, Department de F\'isica i Enginyeria Nuclear,\cr Comte Urgell 187, E-08036 Barcelona, Spain}

\title{Normalizable fermion modes in a holographic superconductor}

\authors{Steven S. Gubser,${}^\PU$ Fabio D. Rocha,${}^\PU$ and Pedro Talavera${}^\Barcelona$}

\abstract{We consider fermions in a zero-temperature superconducting anti-de Sitter domain wall solution and find continuous bands of normal modes.  These bands can be either partially filled or totally empty and gapped.  We present a semi-classical argument which approximately captures the main features of the normal mode spectrum.}

\date{November 2009}

\maketitle

\tableofcontents

\section{Introduction}
\label{INTRODUCTION}

The Abelian Higgs model in anti-de Sitter space \cite{Gubser:2008px,Hartnoll:2008vx} is the most straightforward way to realize superconducting black holes in string theory \cite{Gubser:2009qm,Gauntlett:2009dn,Gubser:2009gp}.  At the level of supergravity calculations, it is a purely bosonic model: the lagrangian involves gravity, an abelian gauge field, and a complex scalar which is the order parameter for superconductivity.  Already in \cite{Gubser:2008px} it was suggested that the complex scalar should be the dual of an operator which destroys a Cooper pair.  However, the dual field theory typically involves both fermions and bosons.  For example, in the construction of \cite{Gubser:2009qm}, the dual of the complex scalar is a sum of a fermion bilinear and a scalar trilinear.  One might imagine a case where the field theory has no scalars (or, at least, none charged under the global symmetry dual to the $U(1)$ gauge symmetry of the bulk that gets spontaneously broken).  Although no such example has been exhibited explicitly, there also isn't any argument we know of that there can't be one.

With fermions in the field theory, one can certainly consider color singlet operators dual to fermions in the bulk: for example, operators schematically of the form $\tr X^n \lambda$ with $n\geq 1$ could be dual to spin-$1/2$ fermions, and $\tr X^n \lambda\lambda\lambda$ with $n\geq 0$ could be dual to either spin-$1/2$ or spin-$3/2$ fermions, depending on how indices are contracted.  (More properly, one must anticipate that bulk fermions are dual to a linear combination of operators with an odd number of field theory fermions.)  It's difficult to have a bulk fermion dual to an operator with one field theory fermion and nothing else, because it's individual fermions in the field theory aren't gauge singlets.  Calculations with fermions \cite{Liu:2009dm,Cubrovic:2009ye,Faulkner:2009wj,Maity:2009zz} have focused on an properties of the two-point function of fermionic operators in the background of a Reissner-Nordstrom-anti-de Sitter black hole (hereafter RNAdS), in particular a singularity at finite momentum which has been argued to be evidence of non-Fermi liquid behavior.

In this paper we wish to study the behavior of spin-$1/2$ bulk fermions in response to the superconducting $AdS_4$ domain wall solution of \cite{Gubser:2009gp}, a zero-temperature geometry whose finite- and low-temperature limits were previously studied in \cite{Gauntlett:2009dn}.  This domain wall solution has two advantages over the four-dimensional RNAdS solution: 1) RNAdS is unstable toward superconducting instabilities, and the domain wall solution is plausibly the endpoint of the evolution of one such instability; 2) RNAdS has macroscopic entropy at zero temperature which has not found a satisfactory explanation, but the domain wall solution has no entropy at all, at least in the classical supergravity approximation.  A technical disadvantage of the domain wall solution is that it is known only numerically, so to study fermions we will have to solve differential equations whose coefficients are known only numerically.  Based on studies including \cite{Gubser:2008wz,Gubser:2008pf,Gubser:2009cg,Horowitz:2009ij}, we expect that domain wall solutions with at least Lorentz invariance in the infrared are fairly generic ground states of the Abelian Higgs model, although Lifshitz scaling in the infrared is another possibility.  Domain walls with conformal invariance in both the ultraviolet (UV) and infrared (IR) are under the best theoretical control, since curvatures can be made small everywhere.  But we anticipate that arguments presented here could be extended to more general domain wall solutions, and the conclusions might be fairly similar when there is emergent Lorentz symmetry in the infrared.

Ideally, the fermions we should study in the $AdS_4$ domain wall  should be the ones present in maximal $d=4$ supergravity (or some consistent truncation thereof).  There are two reasons we do not do this.  The first is simplicity: The quadratic actions for fermions in these theories are complicated, and there appears to be significant mixing between the gravitini and spin-$1/2$ fermions.  The second is flexibility: We will consider different values of the charge and mass of the fermion, and we will see that the final results significantly depend on the choice of these parameters.

The remainder of this paper is structured as follows.  In section~\ref{BACKGROUND} we briefly review the domain wall background.  In section~\ref{PHASE} we present a semi-classical argument which focuses attention on a compact region of phase space and suggests that the normal modes will lie approximately along segments of hyperbolas.  In section~\ref{FOURDIM} we numerically solve the Dirac equation for a charged fermion in the $AdS_4$ domain wall solution, finding one or more bands of fermion normal modes.  We conclude with a discussion in section~\ref{DISCUSSION}.

When this paper was nearing completion, we received \cite{Chen:2009pt}, which studies fermions in a related geometry.  We were also informed by D.~Vegh that he, T.~Faulkner, G.~Horowitz, J.~McGreevy, and M.~Roberts are also working on related topics.

\section{The bosonic background}
\label{BACKGROUND}

\begin{figure}[h]
  \centerline{\includegraphics[width=4in]{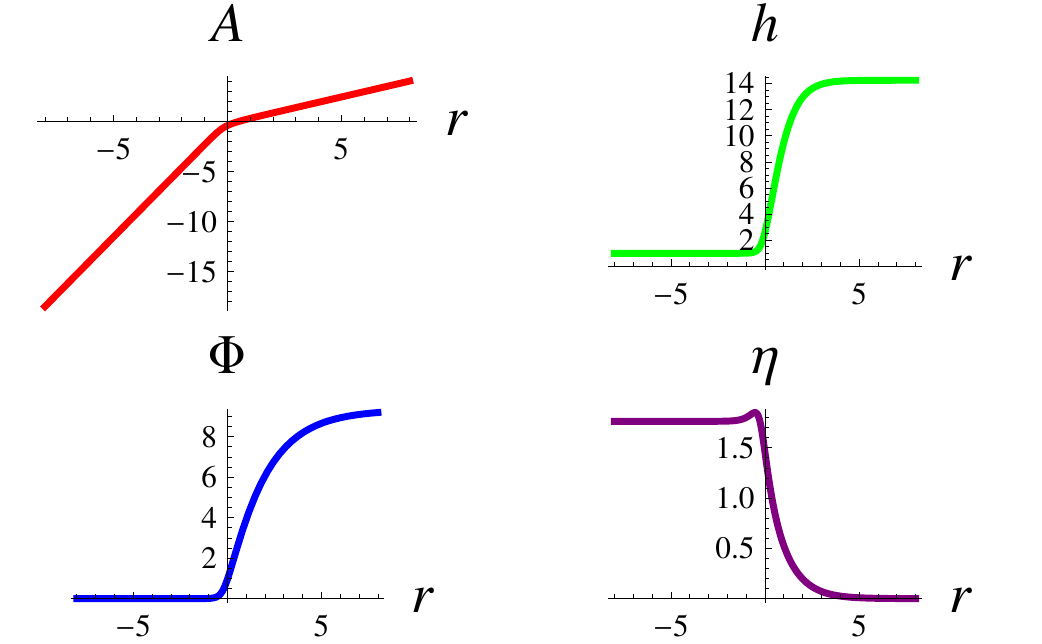}}
  \caption{The metric, gauge field and scalar field for the domain wall solution in M-theory.}\label{fig:domainwall}
 \end{figure}

The bosonic lagrangian that is the basis for the domain wall background that we will study takes the form
 \eqn{Lfour}{
  \qquad& {\cal L} = R - {1 \over 4} F_{\mu\nu}^2 - 
   {1 \over 2} \left[ (\partial_\mu \eta)^2 + \sinh^2 \eta 
    \left( \partial_\mu \theta - {1 \over L} A_\mu \right)^2 \right]
     \cr &\qquad{} 
    + {1 \over L^2} \cosh^2 {\eta \over 2} (7 - \cosh\eta)
 }
where $\eta$ is a real scalar, $\theta$ is a real pseudoscalar, and we use mostly plus metric conventions. We refer the reader to the literature \cite{Gubser:2009gp,Gubser:2008wz} for details on the construction of the domain wall solution, but summarize some of its properties for convenience. The domain wall geometry takes the form
 \eqn{DomainAnsatz}{
  ds^2 = e^{2A(r)} \left[ -h(r) dt^2 + d\vec{x}^2 \right] + 
   {e^{2B(r)} dr^2 \over h(r)} \,,
 }
and has non-zero $\eta(r)$ and gauge field $A_\mu dx^\mu=\Phi(r) dt$. We can always choose coordinates where $B=0$, and will make this choice in numerical computations but will keep $B$ general in formulae. The scalar potential in \eqref{Lfour} has two extrema, at $\eta=0$ and $\eta_{\rm IR}\equiv\log(3+2^{3/2})$, and associated to these are $AdS_4$ solutions with radii of curvature $L_{\rm UV}=L$ and $L_{\rm IR}= \sqrt{3} L/2$, respectively. The domain wall interpolates between these two AdS geometries. We can choose coordinates such that  as $r \to - \infty$
\eqn{IRAdS}{
A \sim {r\over L_{\rm IR}} \qquad h \sim 1 \qquad \eta \sim \eta_{\rm IR} \qquad \Phi \sim 0 \,,
}
while as $r \to +\infty$,
\eqn{UVAdS}{
A \sim {r\over \sqrt{h_{\rm UV}} L_{\rm UV}} \qquad h \sim h_{\rm UV} \qquad \eta \sim 0 \qquad \Phi \sim \Phi_{\rm UV} \,,
}
where $h_{\rm UV}$ and $\Phi_{\rm UV}$ can only be determined numerically and are given by $h_{\rm UV}\approx 14.249$ and $\Phi_{\rm UV}=9.328$. The solution is shown in Fig.~\ref{fig:domainwall}. At a constant $r$ slice the metric is, up to a constant rescaling, the Minkowski metric with an effective ``speed of light" given by $v(r)=\sqrt{h(r)}$. The value of $v(r)$ can be changed by a redefinition of the $t$ and $x$ coordinates, but ratios of $v$ at different values of $r$ are invariant. In our choice of coordinates, $v_{\rm IR}\equiv v(-\infty)=1$ and $v_{\rm UV} \equiv v(+\infty)=3.775$.
 
\section{A semi-classical argument}
\label{PHASE}

The fermionic lagrangian that we will consider is
 \eqn{FermionicL}{
  {\cal L}_f = i \bar\psi (\Gamma^\mu D_\mu - m) \psi \,,
 }
where $D_\mu \psi = \left( \partial_\mu + {1 \over 4} \omega_\mu{}^{\underline{\rho\sigma}} \Gamma_{\underline{\rho\sigma}} - i q A_\mu \right) \psi$ and $\omega_\mu{}^{\underline{\rho\sigma}}$ is the spin connection.  We employ the same conventions on fermions in $AdS_4$ as in \cite{Liu:2009dm}, except that we use $\mu$ for a curved space index and $\underline\mu$ for a tangent space index.  We always assume $m \geq 0$.

Much can be learned from the asymptotics of solutions to the Dirac equation following from \eno{FermionicL} in the UV and IR asymptotic regions.  These asymptotics are simple because all one is describing is a free fermion propagating in empty anti-de Sitter space with a constant electrostatic potential $\Phi$.  Our discussion here will be somewhat heuristic, relying on replacing the Dirac equation by an on-shell relation which is only justified in a geometric optics limit.  We will give a more precise treatment in section~\ref{FOURDIM}.

The on-shell constraint implied by the Dirac equation is
 \eqn{Onshell}{
  g^{\mu\nu} (k_\mu - q A_\mu) (k_\nu - q A_\nu) + m_{\rm eff}^2 = 0 \,.
 }
Here $A_\mu = (\Phi,0,0,0)$ and $k_\mu = (-\omega,k_1,k_2,k_r)$.  Without loss of generality one can use the rotational symmetry in the $x^1$-$x^2$ directions to set $k_2=0$ and $k_1 = k \geq 0$.  The frequency $\omega$ and the momentum $k$ are definite real numbers when we choose the wave-function to take the form $e^{-i\omega t + i k x^1}$ times some function of $r$.  The radial momentum varies as a function of $r$.  The effective mass $m_{\rm eff}^2$ is the bare mass $m^2$ plus some contributions from the spin connection and curvatures.  In components, \eno{Onshell} becomes
 \eqn{OnshellAgain}{
  -{(\omega + q\Phi)^2 \over h} + k^2 + e^{2(A-B)} k_r^2 + 
    e^{2A} m_{\rm eff}^2 = 0 \,.
 }
We next observe that the last term is suppressed in the infrared compared to that first two, unless $m_{\rm eff}^2$ increases quickly in the infrared.  This is impossible if the infrared geometry is anti-de Sitter space: indeed, in that case, $m_{\rm eff}^2$ is constant in the infrared.  In geometries where only Lorentz symmetry is recovered in the infrared, it is possible that $m_{\rm eff}^2$ does increase very quickly, either because of the contributions from the spin connection and curvatures, or from a dependence of $m$ on some scalar field which diverges as one passes to the extreme infrared.  Barring such a circumstance, we see that the sign of $k_r^2$ is the same in the infrared as the sign of $(\omega+q\Phi)^2/h_{\rm IR} - k^2$.  Limiting ourselves now to the case where $\Phi \to 0$, we see that $k_r^2$ is non-positive when
 \eqn{IRcondition}{
  {|\omega| \over v_{\rm IR}} \leq k \qquad\hbox{where}\qquad v_{\rm IR} = \sqrt{h_{\rm IR}} \,.
 }
Now, if $k_r$ were real an non-zero, then its value would be essentially the radial momentum of the fermion.  As an extension of standard boundary conditions at a black hole horizon, one may reasonably require this momentum to be infalling (that is, toward the infrared).  But it makes no sense for the fermion to be falling down ever further into the infrared when its state is a normal mode.  Instead, $k_r$ should be either $0$ or imaginary, so that the wave-function of the fermion can decay as one passes further into the infrared.  If the infrared geometry is anti-de Sitter, then the decay is very rapid: exponential in $|k_r| e^{-A}$, or power-like in $e^{-A}$ if $k_r = 0$.  In a more general setting, the right requirement is for the fermion wave-function to be normalizable in the infrared, which presumably rules out oscillatory and singular behavior while allowing a more regular solution similar to the exponentially damped solution in anti-de Sitter space.  The upshot of this discussion is that \eno{IRcondition} is one requirement that a normal mode must satisfy.  

The requirement that $\omega=0$ for a fermionic normal mode in RNAdS comes from reasoning rather similar to what we just went through: only for this value of $\omega$ can one avoid oscillatory behavior in the infrared indicating that the fermion is falling into the black hole.  In fact, since $h \to 0$ at the horizon of the RNAdS geometry, one has $v_{\rm IR} = 0$, so one can still use $\omega \leq v_{\rm IR} |k|$ as the condition that restricts the possible values of $\omega$ and $k$ based on the infrared dynamics.

In the extreme ultraviolet, the last term of \eno{OnshellAgain} dominates over the first two, so one can conclude that $k_r^2$ has the sign opposite $m_{\rm eff}^2$, namely negative.  So oscillatory behavior is impossible.  Though our reasoning here is non-rigorous, the conclusion that solutions are non-oscillatory is correct: all components of the fermion wave-function must have power-law behavior in $e^{-A}$, no matter what $\omega$ and $k$ are.  Now, it is hard to see how one would find a normal mode if $k_r^2$ were always negative, because then either the wave-function of the fermion would tend to increase monotonically toward the ultraviolet or else monotonically toward the infrared, neither of which would be consistent with normalizable behavior.  So, as a necessary condition beyond \eno{IRcondition}, it is reasonable to expect that 
 \eqn{supCondition}{
  \sup_r k_r^2 = \sup_r \left\{ {(\omega+q\Phi)^2 \over h} - k^2 - e^{2A} m_{\rm eff}^2 
     \right\} > 0 \,.
 }
In principle, knowing a bosonic background, one can use \eno{supCondition} to obtain an upper bound on values of $k$ where normal modes can exist, as a function of $\omega$.  In practice this is laborious.  We prefer to use instead the related condition:
 \eqn{UVCondition}{
  {|\omega+q\Phi_{\rm UV}| \over v_{\rm UV}} > k \,,
 }
where $v_{\rm UV} = \sqrt{h_{\rm UV}}$.  \eno{supCondition} and \eno{UVCondition} are equivalent in the limit where $\omega/v_{\rm UV}$ and $k$ are large compared to $m_{\rm eff}$.  So \eno{UVCondition}, which is simple to apply, is really the large frequency, large wave-number limit of the better justified condition \eno{supCondition}.

\begin{figure}[h!]
  \centerline{\includegraphics[width=4in]{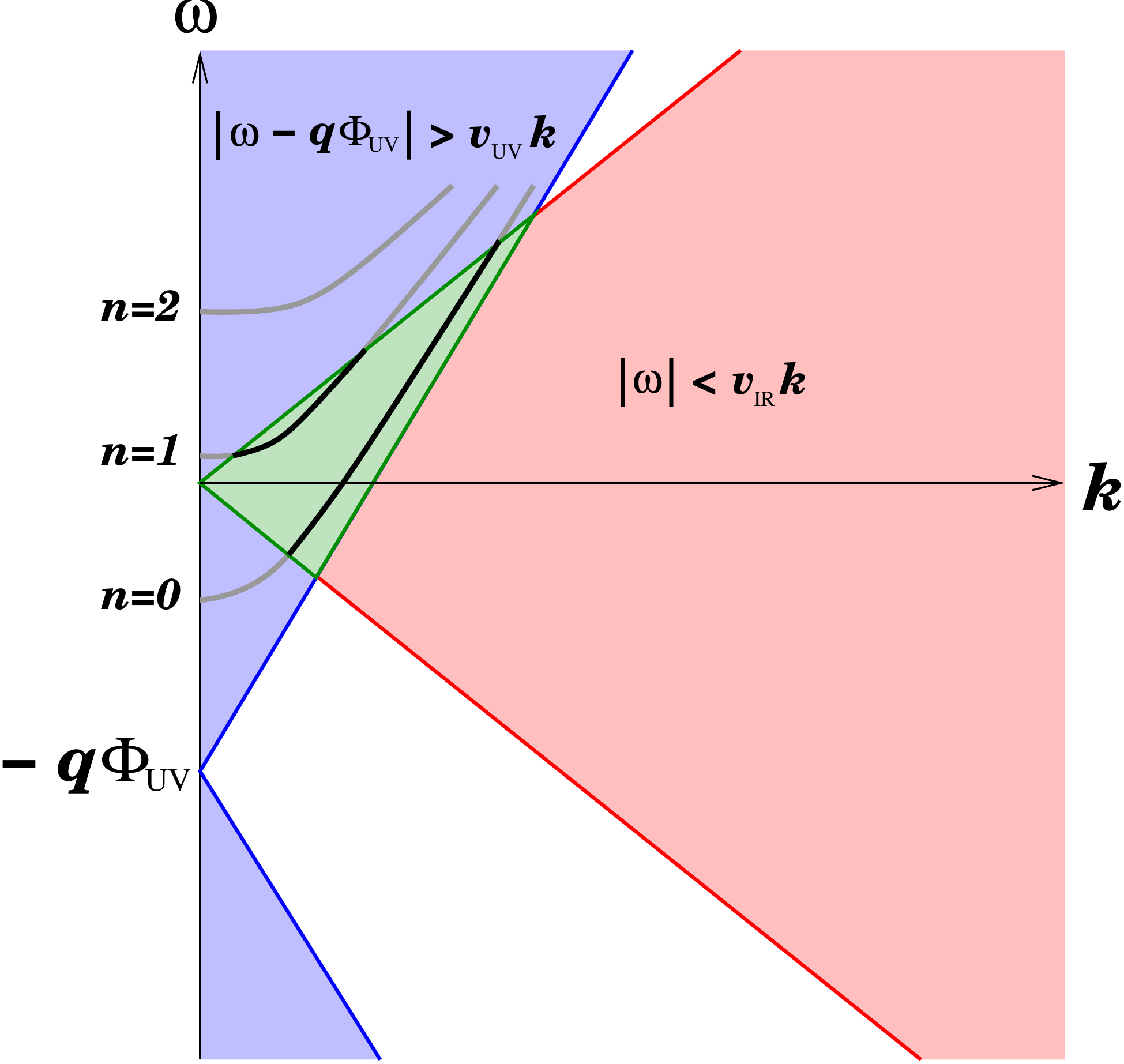}}
  \caption{The green wedge is where \eno{SummaryRelation} holds.  This is the region where our heuristic geometric optics arguments indicate that one might find fermion normal modes.  The grey and black curves are an approximate depiction of the hyperbolas \eno{HyperboloidEstimate}, representing an approximate WKB treatment of where normal modes lie.  Only the black parts of the curve correspond to actual normal modes; the grey parts are where normal modes might have been if the green wedge had been larger.}\label{Wedge}
 \end{figure}

To summarize: A heuristic and inexact treatment of asymptotics in the UV and IR regions of a domain wall geometry leads to the expectations that normal modes will be confined to what we will term the ``preferred wedge,'' namely
 \eqn{SummaryRelation}{
  {|\omega| \over v_{\rm IR}} \leq k < {|\omega + q\Phi| \over v_{\rm UV}} \,.
 }
In other words, the projection of the gauge-invariant momentum $k_\mu - q A_\mu$ onto boundary directions should be timelike in the ultraviolet, in order to produce non-monotonic dependence on $r$, and spacelike in the infrared, in order to ensure that the fermions aren't falling into the far infrared.  The first inequality in \eno{SummaryRelation} (from the infrared constraint) can be expected to be more reliable than the second inequality (from the ultraviolet behavior).  The preferred wedge is compact, as figure~\ref{Wedge} shows.

Having established our expectations for the region of the $\omega$-$k$ plane where normal modes may appear, the next obvious question is what the pattern of normal modes will be in this region.  First, if there are normal modes at all, it is reasonable to think that they form one-dimensional families.  The reason is that individual components of the Dirac spinor satisfy a second-order differential equation, so there are two linearly independent solutions.  Demanding that the one that is normalizable in the infrared is also normalizable in the ultraviolet amounts to a single real condition on the two real parameters $\omega$ and $k$.  So indeed one expects one-dimensional families.

A Bohr-Sommerfeld estimate of the location of normal modes takes the schematic form
 \eqn{BohrSest}{
  \int\limits_{r_{\rm IR}(\omega,k)}^{r_{\rm UV}(\omega,k)} dr \, k_r = 
   \pi (n + \nu) \,.
 }
Here $r_{\rm UV}$ and $r_{\rm IR}$ are the radii where $k_r=0$ is a solution to \eno{OnshellAgain}; $k_r$ is the positive root of \eno{OnshellAgain} for $r_{\rm IR} < r < r_{\rm UV}$; $n$ is a non-positive integer; and a standard expectation for $\nu$ (the Keller-Maslov index) is $1/2$.  Passing to a further approximation, one can consider the case where the integral \eno{BohrSest} is dominated by radii where the geometry is approximately $AdS_{\rm UV}$ and the gauge potential is approximately $\Phi_{\rm UV}$.  Then the integral \eno{BohrSest} should be roughly proportional to $\sqrt{{(\omega+q\Phi)^2 \over h} - k^2}$, with a constant of proportionality that depends on details of the bosonic background and so cannot readily be computed.  Using \eno{BohrSest}, we are led to an approximate condition
 \eqn{HyperboloidEstimate}{
  {(\omega+q\Phi_{\rm UV})^2 \over v_{\rm UV}^2} - k^2 = (Y_1 n + Y_2)^2
 }
for the position of the $n^{\rm th}$ band of normal modes.  $Y_1$ and $Y_2$ are constants and $n$ is again a non-positive integer, which is the number of nodes in the fermion wave-function.  (Different components of the Dirac spinor could have bands of normal modes described by \eno{HyperboloidEstimate} with different $Y_2$ but the same $Y_1$.  Associating $n$ with the number of nodes can only be expected to work when one is focusing on the normal modes involving a particular component.)  Evidently, we should expect only finitely many bands of normal modes, because the hyperbola \eno{HyperboloidEstimate} doesn't intersect the region $|\omega| < v_{\rm IR} k$ when $n$ is large.

Since the hyperbolas \eno{HyperboloidEstimate} do not intersect the UV boundary $|\omega-q\Phi_{\rm UV}| = v_{\rm UV} k$ of the preferred wedge region for normal modes, the topological prediction we get out is that the bands begin and end on the IR boundary, $|\omega| = v_{\rm IR} k$.  It is possible for a band to both begin and end on the upper branch of the IR boundary, i.e.~the one with $\omega>0$.  Herein lies an interesting possibility: for such a band, there would be a minimum energy required to add a fermion belonging to it (unless the band dipped below the $\omega=0$ axis and then rose back up).  This is obviously an attractive feature if one is to make contact between these holographic models of superconductors and real-world superconductors.  However, we are obliged to add that it is equally possible for bands of normal modes to cross the $\omega=0$ axis one or more times.  Then the natural configuration to consider is the one where the $\omega < 0$ normal modes are populated, while the $\omega>0$ modes aren't.  This results in a Fermi surface in the bulk, and apparently a gapless state for the fermions.  Two considerations might change this state of affairs: 1) The fermions interact gravitationally and so presumably have an attractive channel that causes them to superconduct in the bulk through the usual mechanism of forming Cooper pairs; and 2) The gas of fermions back-reacts on the bulk, as in \cite{deBoer:2009wk}.

In the next section, we will exhibit a case where most bands cross the $\omega=0$ surface at least once, and another case where there is a single band that does not cross the $\omega=0$ axis and therefore has a gap in the sense explained in the previous paragraph.  It is interesting indeed to inquire what behavior the fermions of maximal gauged supergravity exhibit in the domain wall geometry summarized in section~\ref{BACKGROUND}: gapped or ungapped?

As a final aside, it is easy to imagine cases where there are no normal modes: the condition \eno{OnshellAgain} could be impossible to satisfy, or it could be that $Y_2$ in \eno{HyperboloidEstimate} is too large for any of the would-be normal modes to intersect with permitted infrared boundary conditions.  This doesn't seem quite to be a situation meriting the term ``gapped behavior,'' because there isn't a definite energy that one can add to the system to obtain a long-lived fermionic quasi-particle.  But perhaps it is reasonable to expect that in a strongly interacting superconductor at zero temperature, unpaired fermions can never be stable.  Normal modes could be altogether absent in ground states with emergent conformal or Lorentz symmetry, where the preferred wedge is (modulo caveats already discussed) really a triangular wedge as shown in figure~\ref{Wedge}; or they could be altogether absent in ground states with Lifshitz symmetry, where $v_{\rm IR}=0$ and the preferred wedge shrinks to a line segment along the $\omega=0$ axis; or, indeed, they could be altogether absent in an RNAdS vacuum: this happens if $q$ is too small.  It would be interesting to study the spectral measure of the fermion two-point function in situations where there are no normal modes.  It could be that a ridge similar to the one found, for example, in \cite{Liu:2009dm}, would remain, suggestive of unstable fermionic quasi-particles.

\section{Bands of fermion normal modes in $AdS_4$}
\label{FOURDIM}

Let us now describe the numerical computation of the normal modes in more detail. It is convenient to choose a basis of $\Gamma$ matrices such that $\Gamma^{\underline{r}}$ is diagonal, such as
\eqn{Gammadef}{
\Gamma^{\underline{r}} = \begin{pmatrix} 1 &0 \\ 0& -1\end{pmatrix} \,,\qquad \Gamma^{\underline{i}}=\begin{pmatrix} 0&\gamma^i \\ \gamma^i  & 0 \end{pmatrix} \qquad i=0,1,2 \,,
}
where $\gamma^i$ are gamma matrices in three dimensions given by $\gamma^0=i\sigma_2$, $\gamma^1=\sigma_1$, and $\gamma^2=\sigma_3$. As argued above, symmetry allows us to take the spinor to be of the following form:
\eqn{Spinorcomp}{
\psi(t,\vec{x},r) = e^{-i \omega t + i k x^1} u(r)  =e^{-i\omega t + i k x^1} 
  \begin{pmatrix} u^+_1(r) \\ u^+_2(r) \\ u^-_1(r) \\ u^-_2(r) \end{pmatrix} \,.
}
The Dirac equation can then be written as
\eqn{Diracb}{
\left[ \sqrt{h} \Gamma^{\underline{r}} \partial_r +  i e^{-A}\left(k \Gamma^{\underline{1}} - \frac{\omega +q \Phi}{\sqrt h} \Gamma^{\underline{0}}\right) +\frac{6 h A' +h'}{4\sqrt h}\Gamma^{\underline{r}} - m\right] u =0 \,.
}

Let us now consider the asymptotic behavior of solutions to \eqref{Diracb}. We are interested in regular and purely infalling solutions, which implies that in the infrared (corresponding to $r\to-\infty$), 
\eqn{SpinorIR}{
u^\pm_a(r) \approx U^\pm_a  e^{-2A}K_{{1\over 2} \pm m L_{\rm IR}} \bigg( \kappa_{\rm IR} e^{-A}\bigg) \qquad \kappa_{\rm IR} \equiv L_{\rm IR}\sqrt{k_1^2-\omega^2} \,.
}
Here $a=1,2$, $K_\nu$ is a modified Bessel function, and the $U^\pm_a$ are constants that satisfy
\eqn{Uirel}{
\frac{U^-_2}{U^+_1} =  -\frac{U^+_2}{U^-_1}= i \sqrt{\frac{k_1+\omega}{k_1-\omega}}
}
and are otherwise arbitrary. In what follows, we will set $U^+_1=U^+_2=1$.

In the ultraviolet, corresponding to $r\to+\infty$, the most general solution is of the form
\eqn{uInAdS4}{
u^{\pm}_a = C^{\pm}_a e^{-2 A}  I_{\mp{1\over 2} -m L_{\rm UV}}\Big(\kappa_{\rm UV} e^{-A}\Big) + D^{\pm}_a e^{-2 A}  I_{\pm{1\over 2} +m L_{\rm UV}}\Big(\kappa_{\rm UV} e^{-A}\Big)\,,
}
where $\kappa_{\rm UV}^2 \equiv k_1^2-\frac{(\omega+q\Phi_{\rm UV})^2}{h_{\rm UV}}$ and $I_\nu$ is a modified Bessel function. As before, \eqref{Diracb} allows you to solve for $C^-_a,D^-_a$ algebraically in terms of $C^+_a, D^+_a$, but the exact form of this relation will not be important.

Expanding \eqref{uInAdS4} at large $r$, we see that
\eqn{uLarger}{
u^+_a \approx C^+_a \left(\kappa_{\rm UV} \over 2\right)^{-{1\over 2} - m L_{\rm UV}}{e^{\left(-{3\over2}+m L_{\rm UV}\right)A}\over\Gamma\left({1\over 2} -m L_{\rm UV} \right)}+ 
D^+_a  \left(\kappa_{\rm UV} \over 2\right)^{{1\over 2} + m L_{\rm UV}}{e^{\left(-{5\over2}-m L_{\rm UV}\right)A}\over\Gamma\left({3\over 2} -m L_{\rm UV} \right)} \,.
}
According to the standard AdS/CFT prescription, we identify $C^+_a$ with the source and $D^+_a$ with the response of a boundary fermionic operator with conformal dimension $\Delta={3\over2} + m L_{\rm UV}$.\footnote{For $0\leq m < 1/2$, it is also legitimate to identify $D^+_a$ with the source and $C^+_a$ with the response, but we will in this paper stick with the identification valid for all $m$.} 

Normal modes are then solutions of \eqref{Diracb} that in the infrared satisfy \eqref{SpinorIR} and have $C^+_a=0$. They correspond to poles of the retarded Green's function. To make this connection more explicit, consider the full fermionic action
 \eqn{FermionicS}{
  S_f = i\int d^5 x\,\bar\psi (\Gamma^\mu D_\mu - m) \psi +S_{\rm bdy}\,
 }
 where $S_{\rm bdy}$, which we neglected up to now, is a boundary term that does not contribute to the equations of motion. It is necessary to have a well posed variational problem and gives the only nonzero contribution to the onshell action. We choose
 \eqn{Sbdydef}{
S_{\rm bdy} = -i \int\limits_{r=1/\varepsilon} d^4 x\, \sqrt{-g g^{rr}} \, \bar\psi_+ \psi_-\,, \qquad \psi_\pm\equiv {1\over 2} \left(1\pm\Gamma^{\underline r}\right)\psi \,,
}
where $\varepsilon$ is a positive quantity to be taken to zero after functional derivatives are taken. By taking the appropriate functional derivatives of \eqref{FermionicS}, we obtain the retarded Green's function \cite{Iqbal:2009fd}
\eqn{GotGR}{
G_R = \left(\kappa_{\rm UV} \over 2\right)^{mL_{\rm UV}} {\Gamma\left({1\over2} - m L_{\rm UV}\right)\over\Gamma\left({1\over2} + m L_{\rm UV}\right)} \begin{pmatrix} - i \frac{D^-_1}{C^+_2} &0 \\ 0 & i \frac{D^-_2}{C^+_1} \end{pmatrix} \,,
}
and we see that zeros of $C^+_2$ are poles of $G_{11}$ while zeros of $C^+_1$ are poles of $G_{22}$. From \eqref{Diracb} and the form of the $\Gamma$ matrices \eqref{Gammadef} it is clear that the four components of the spinor $u$ couple to each other only in pairs: $u_1^+$ and $u_2^-$ mix, and so do $u_2^+$ and $u_1^-$. This means we are free to consider only $u^+_1$ and $u^-_2$  nonzero and look for zeros of $C^+_1$  or consider only  $u^+_2$ and $u^-_1$ nonzero and look for zeros of $C^+_2$.

Before discussing our numerical results let us comment on the meaning of the ``preferred wedge" \eqref{SummaryRelation}. It is clear that \eqref{IRcondition} translates\footnote{Remember we are using units where $v_{\rm IR}=1$.}  to $\kappa_{\rm IR}^2>0$ which implies that $u$ is \emph{not} oscillatory in the IR. Condition \eqref{UVCondition} on the other hand, translates to $\kappa_{\rm UV}^2<0$ which implies that $u$ \emph{is} oscillatory in the UV.  This is in agreement with the geometrical optics arguments that led us to formulate these conditions in section~\ref{PHASE}.

We now have the necessary ingredients to find normal modes numerically: we can solve \eqref{Diracb} with initial conditions given by \eqref{SpinorIR} for some large negative $r$. We can then determine the $C^+_a$ coefficients by fitting the numerical results to \eqref{uInAdS4} for large positive $r$ and vary the parameters until we find zeros of  the $C^+_a$. As expected, we find a continuous set of normal modes inside the ``preferred wedge". As you increase $q$, the ``preferred wedge" grows and accommodates more bands. 

\begin{figure}[h!]
\begin{center}
 \includegraphics[width=6in]{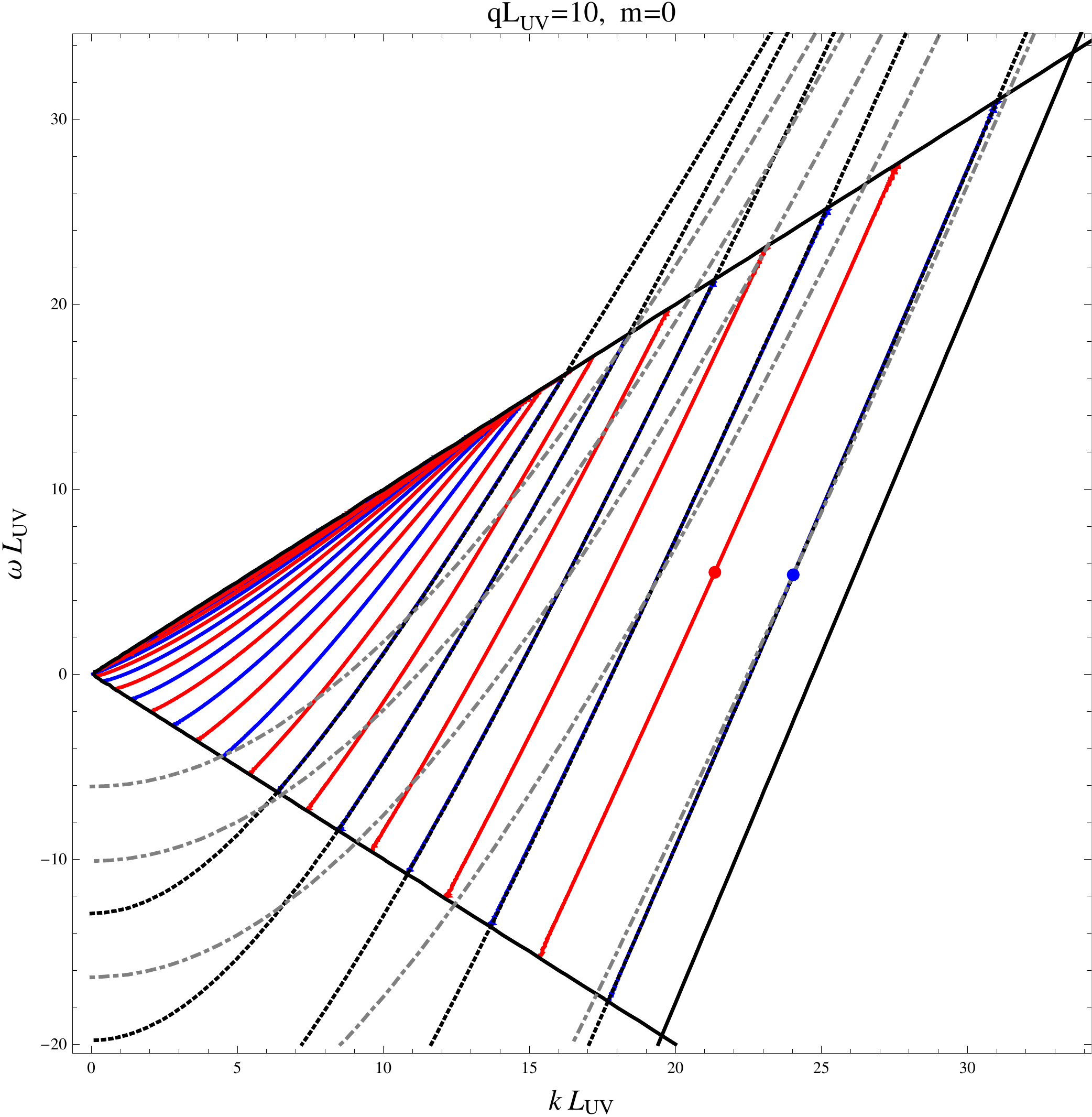}
\end{center}
\caption{
Fermion normal modes in the $AdS_4$ domain wall for $m=0$ and $q L_{\rm UV}=10$. The black lines mark the boundary of the ``preferred wedge" \eqref{SummaryRelation}. The red lines correspond to normal modes where $u^+_1$ and $u^-_2$ are nonzero (poles in $G_{22}$) while the blue lines correspond to normal modes where the $u^+_2$ and $u^-_1$  are nonzero (poles in $G_{11}$). The gray dot-dashed lines are one parameter fits and the black dashed lines are three parameter fits to hyperbola. The red and blue dots mark the location of the normal modes shown in Fig.~\ref{fig:twomodes}.
\label{fig:normalmodesqL10}}
\end{figure}

\begin{figure}[h!]
\begin{center}
 \includegraphics[width=6.5in]{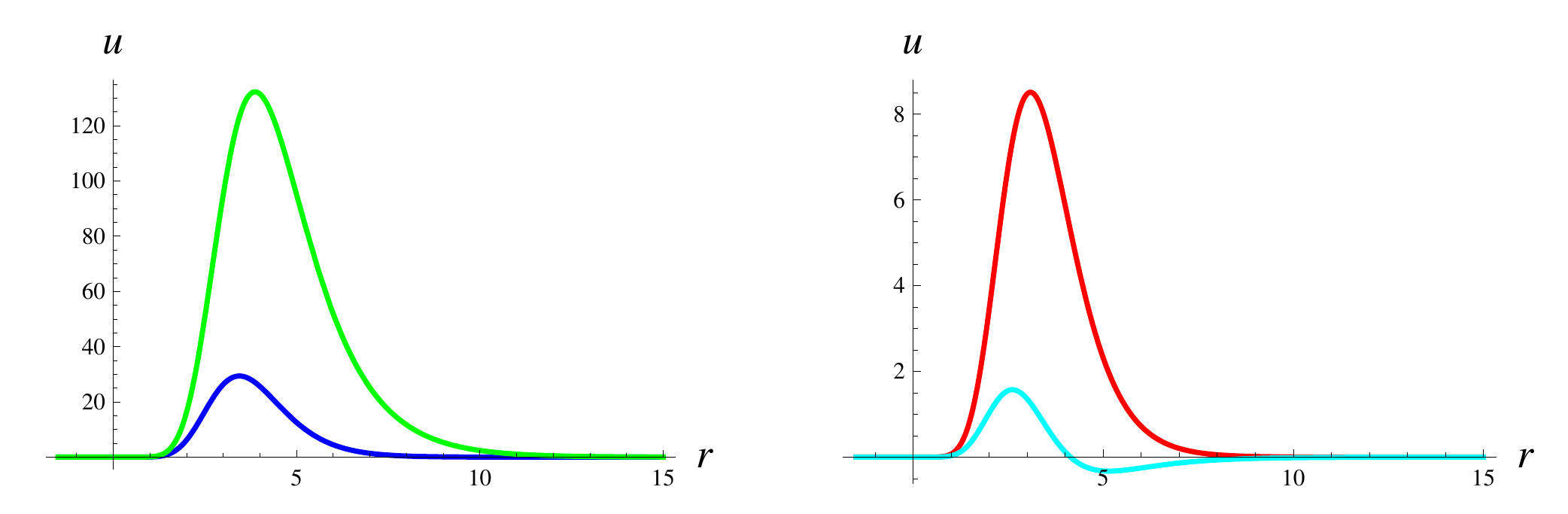}
\end{center}
\caption{
The wave functions for two fermion normal modes of the $AdS_4$ domain wall for $m=0$ and $q L_{\rm UV}=10$. We note that in the $\kappa_{\rm IR}>0$ region, we can always choose the $u^+_a$ to be purely real, in which case it follows that the $u^-_a$ are purely imaginary. The plot on the left shows the real part of $u^+_1$ (blue) and the imaginary part of $u^-_2$ (green) corresponding to the blue dot in Fig.~\ref{fig:normalmodesqL10}. The plot on the right shows the real part of $u^+_2$ (red) and the imaginary part of $u^-_1$ (cyan) corresponding to the red dot in Fig.~\ref{fig:normalmodesqL10}.
\label{fig:twomodes}}
\end{figure}

As a specific example consider $m=0$ and $q L_{\rm UV}=10$, for which the bands are shown in Fig.~\ref{fig:normalmodesqL10}. We find an abundance of normal modes, with poles of $G_{22}$ alternating with poles of $G_{11}$. The number of nodes in the wave functions (some of which are shown in Fig.~\ref{fig:twomodes}) is greater for bands that are closer to the origin. The bands are well approximated by hyperbola, although \eqref{HyperboloidEstimate} does not seem to capture their shape very well. Generalizing \eqref{HyperboloidEstimate}, we can fit the bands to
 \eqn{HyperboloidFit}{
  {(\omega+q\Phi)^2 \over v^2} - k^2 = m_{\rm eff}^2\,,
 }
where we treat $m_{\rm eff}$ as an arbitrary fitting parameter. We can either use the extreme UV values of $\Phi$ and $v$ and do one-parameter fits (shown as gray dot-dashed lines in Fig~\ref{fig:normalmodesqL10}) or treat $\Phi$ and $v$ as arbitrary fitting parameters also and do three-parameter fits (shown as black dashed lines in Fig~\ref{fig:normalmodesqL10}). In the former case, we find that $m_{\rm eff}$ increases as a small power of the band number.  In the latter case (unsurprisingly) we obtain better fits, and we note that the best fit values of $\Phi$ and $v$ approach the expected UV values when you consider bands farther from the origin.

\begin{figure}[h!]
\begin{center}
 \includegraphics[width=3in]{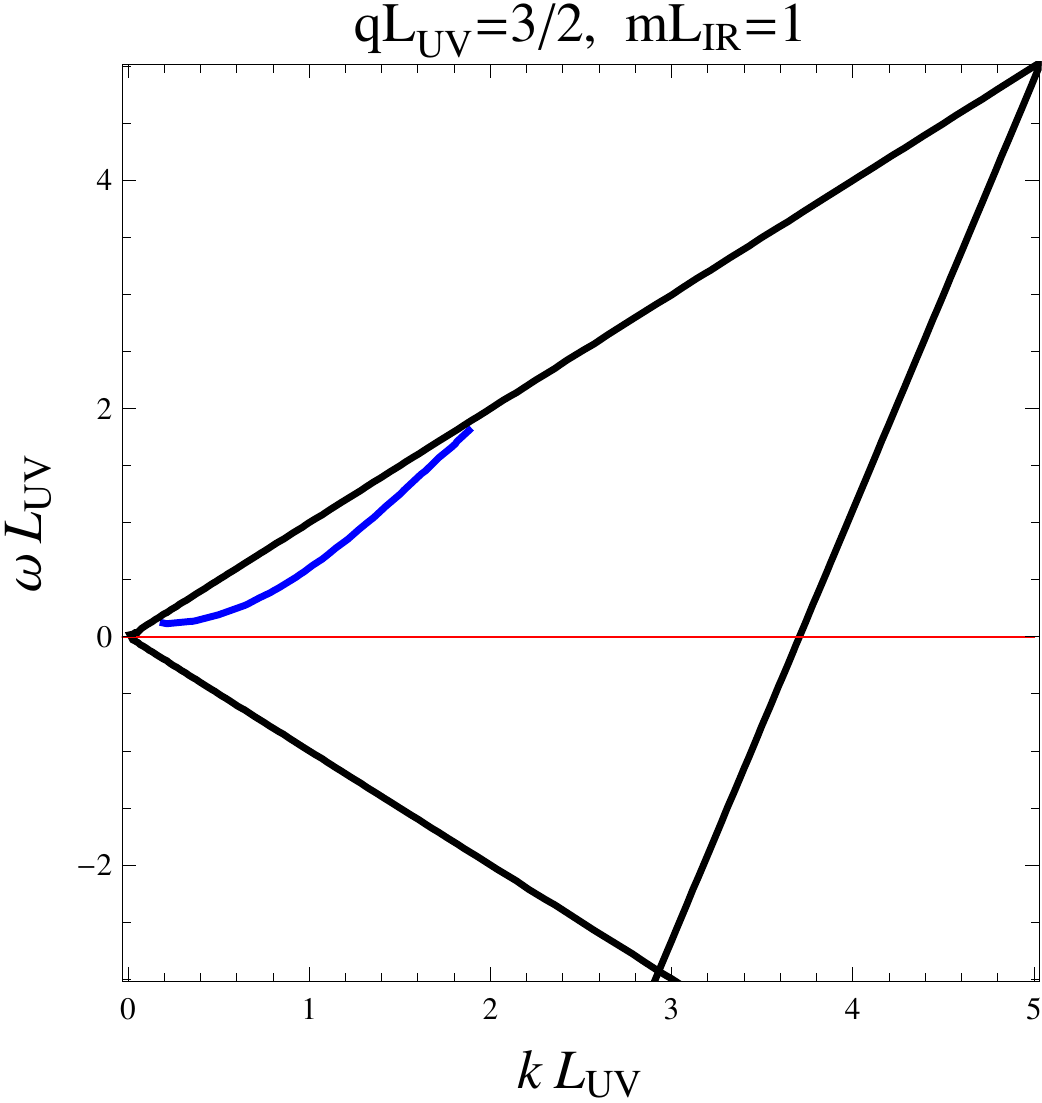}
\end{center}
\caption{
Fermion normal modes of the $AdS_4$ domain wall for $m L_{\rm IR}=1$ and $q L_{\rm UV}=3/2$. The black lines mark the boundary of the ``preferred wedge" \eqref{SummaryRelation}. The blue line corresponds to normal modes where the $u^+_2$ and $u^-_1$  are nonzero (poles in $G_{11}$). Notice it never intersects the $\omega=0$ line (red).  So this is a gapped band.
\label{fig:gappedband}}
\end{figure}

In Fig.~\ref{fig:normalmodesqL10}, all the bands cross the $\omega=0$ line and are therefore ungapped. As you decrease $q$, the ``preferred wedge" shrinks and fewer bands are present, but for $m=0$ at least one of them always seems to intersect the $\omega=0$ line. We can argue that this is always the case if we note that for small enough $\omega$ and $k$, the Green's function should approach its AdS value \cite{Iqbal:2009fd}
\eqn{GRAdS}{
G_R = f\left(m L_{\rm IR}\right) \kappa^{2 m L_{\rm IR} -1} \gamma\cdot k \propto \kappa^{2 m L_{\rm IR}} \begin{pmatrix} - \sqrt{k-\omega\over k+\omega} & 0\\ 0 &\sqrt{k+\omega\over k-\omega} \end{pmatrix} \,,
}
where $\kappa^2= k^2-\omega^2$ and $f(m L_{\rm IR})$ is some unimportant constant. Thus, for $m=0$, there is always a divergence at the origin that signals a normal mode with $(\omega,k)=0$ and at least one of the bands is ungapped. For  positive $m$ however, this divergence is gone and we expect the possibility of gapped bands. In fact, for $q L_{\rm UV}=3/2$ and $m L_{\rm IR}=1$, we find a single gapped band (see Fig.~\ref{fig:gappedband}). 

Finally, we note that it is also possible to choose the parameters in such a way that there are no normal modes at all. If $q L_{\rm UV}$ is small enough, this seems to be the generic behavior. For instance, taking $m=0$ and $q L_{\rm UV}=1/8$, we find no normal modes.

\section{Discussion}
\label{DISCUSSION}

We studied charged Dirac fermions in an $AdS_4$ domain wall solution and found that generically they exhibit continuous bands of normal modes. This is in stark contrast to fermions in RNAdS, which exhibit one or more isolated normal modes at $\omega=0$ and finite $k$ \cite{Liu:2009dm,Cubrovic:2009ye,Faulkner:2009wj,Maity:2009zz}. However, we note that the same semiclassical argument that led us to expect continuous bands in the domain wall geometry can be used to argue for isolated normal modes in RNAdS. In fact, since RNAdS has an horizon at which the metric function $h$ vanishes, it can in some sense be thought of as degenerate domain wall for which $v_{\rm IR}=0$. The condition on the location of the normal modes given by\eqref{IRcondition} then degenerates to $\omega=0$ and, choosing units where $v_{\rm UV}=1$,  the remaining condition \eqref{UVCondition} becomes $|q\Phi_{\rm UV}|>k$. The ``preferred wedge'' thus becomes a ``preferred segment'' and the same parameter counting that led us to expect continuous bands for the domain wall geometry tells us that we should find a discrete set of normal modes at $\omega=0$, which is precisely what happens for RNAdS.

Since in our conventions, energy equal to the Fermi energy is equivalent to frequency equal to zero, the natural zero temperature configuration is one where the normal modes with $\omega\leq 0$ are occupied and those with $\omega>0$ are unoccupied. Therefore, a particularly important feature of the bands is whether they cross the $\omega=0$ axis. We found this feature depends on the choice of the charge and mass of the fermion. For zero mass fermions, the bands seem to always cross the $\omega=0$ axis and the resulting configuration would seem to have a Fermi surface. However, to better understand this Fermi surface it would be necessary to consider the back-reaction of the fermions on the bulk geometry, which might conceivably destroy the Fermi surface. For positive mass fermions, with a suitable choice of the charge, we found an example of a single band that does not touch the $\omega=0$ axis, i.e., the band is gapped.  Such a feature is desirable in making comparisons with real-world superconductors at zero temperature, and it may be fairly generally achievable when there is at least emergent Lorentz symmetry in the infrared. 

Because our bosonic background is embeddable in M-theory \cite{Gauntlett:2009dn,Gubser:2009gp}, it would be interesting to redo our calculations using the quadratic fermion action of maximal gauged $d=4$ supergravity.  It appears to be non-trivial to diagonalize this quadratic action, but the advantage is that one should in principle be able to understand the operators dual to the fermions.  It would also be interesting to redo our calculations in $AdS_5$, where we expect qualitatively similar results, simply because the semi-classical arguments of section~\ref{PHASE} don't depend on dimension.  Preliminary numerical studies yielded results in agreement with these expectations.

\section*{Acknowledgments}

FDR would like to thank A. Nellore and S. Pufu for useful discussions. This work was supported in part in part by the DOE under Grant No.\ DE-FG02-91ER40671 and by the NSF under award number PHY-0652782. FDR was also supported in part by the FCT grant SFRH/BD/30374/2006.  The work of P.\ T.\ is
partially supported by FPA2007-66665, Consolider CPAN grant SCD2007-00042, Grup Consolidat 2005SGR00564 and by a MEC PR2008-0332 grant.

\bibliographystyle{ssg}
\bibliography{fermion}

\end{document}